\documentclass[]{aastex631}

\shorttitle{SFMS of GLSBGs}
\shortauthors{Du et al.}
\graphicspath{{./}{figures/}}

\begin{document}
\title{Star-forming Main Sequence of Giant Low Surface Brightness Galaxies}

\author[0000-0003-4546-8216]{Wei Du}
\affiliation{Key Laboratory of Optical Astronomy, National Astronomical Observatories, Chinese Academy of Sciences, Beijing,100101,China; wdu@nao.cas.cn}

\author[0000-0003-0202-0534]{Cheng Cheng}
\affiliation{Chinese Academy of Sciences South America Center for Astronomy, National Astronomical Observatories, CAS, Beijing 100101, China}
\affiliation{Key Laboratory of Optical Astronomy, National Astronomical Observatories, Chinese Academy of Sciences, Beijing,100101,China}

\author{Pengliang Du}
\affiliation{Key Laboratory of Optical Astronomy, National Astronomical Observatories, Chinese Academy of Sciences, Beijing,100101,China}
\affiliation{School of Astronomy and Space Science, University of Chinese Academy of Sciences, 19A Yuquan Road, Shijingshan District, Beijing, 100049, China}

\author{Lin Du}
\affiliation{Key Laboratory of Optical Astronomy, National Astronomical Observatories, Chinese Academy of Sciences, Beijing,100101,China}
\affiliation{School of Astronomy and Space Science, University of Chinese Academy of Sciences, 19A Yuquan Road, Shijingshan District, Beijing, 100049, China}

\author{Hong Wu}
\affiliation{Key Laboratory of Optical Astronomy, National Astronomical Observatories, Chinese Academy of Sciences, Beijing,100101,China}








\begin{abstract}
 Giant Low Surface Brightness Galaxies (GLSBGs) are fundamentally distinct from normal galaxies and other types of low surface brightness galaxies (LSBGs) in star formation and evolution. In this work, we collected 27 local GLSBGs from the literature. They have high stellar masses (M$_{*}>$10$^{10}$ M$_{\odot}$) and low FUV-based star formation rates (SFRs). With the specific SFRs (sSFR) lower than the characteristic value of the local star-forming galaxies of M$_{*}$=10$^{10}$ M$_{\odot}$(sSFR $<$ 0.1 Gyr$^{-1}$),  GLSBGs deviate from the star formation main sequence (MS) defined for local star-forming galaxies respectively by  \citet{2007A&A...468...33E} (E07) and \citet{2016MNRAS.462.1749S} (S16) at the high M$_{*}$ regime. They are H{\sc{i}}-rich systems with the H{\sc{i}} gas mass fractions (f$_{\rm HI}$) higher than the S16 MS galaxies, but have little molecular gas (H$_{2}$), implying a quite low efficiency of H{\sc{i}}-to-H$_{2}$ transition due to the low H{\sc{i}} surface densities ($\Sigma$ H{\sc{i}}) that are far lower than the minimum $\Sigma$ H{\sc{i}} of 6 - 8 M$_{\odot}$ pc$^{-2}$ required for shielding the formed H$_{2}$ from photodissociation.
 For GLSBGs, the inner, bulge-dominated part with lower SFRs and higher M$_{*}$ is the main force pulling the entire GLSBG off from the MS, while the outer, disk-dominated part with relatively higher SFRs and lower M$_{*}$ reduces the deviations from the MS. For some cases, the outer, disk-dominated parts even tend to follow the MS. In the aspect of NUV - $r$ versus $g$ - $r$ colors, the outer, disk-dominated parts are blue and behave similarly to the normal star-forming galaxies while the inner, bulge-dominated parts are in statistics red,  indicating an inside-out star formation mechanism for the GLSBGs. They show few signs of external interactions in morphology, excluding the recent major merger scenario.
\end{abstract}


\section{Introduction} \label{sec:intro}
The star formation and evolution of galaxies are regulated by various environmental-dependent processes, such as mergers and interactions \citep{1977egsp.conf..401T} that could trigger the violent star formation activity,  and strangulation or starvation\citep{1980ApJ...237..692L}, ram pressure stripping\citep{1972ApJ...176....1G}, and galaxy harassment\citep{1996Natur.379..613M} that could quench the star formation of galaxies. Different processes shape galaxies to follow some physical scaling relations, which retain the imprint of galaxy formation and evolution. 
Among those scaling relations, an important one is the relation between star formation rate (SFR) and stellar mass (M$_{*}$) for star-forming galaxies, dubbed as star formation main sequence (MS; \citet{2004MNRAS.351.1151B,2007ApJ...670..156D,2007A&A...468...33E,2007ApJ...660L..47N,2007ApJ...660L..43N,2011ApJ...742...96W}). The MS tells that star-forming galaxies populate a narrow sequence in SFR - M$_{*}$ plane with a small intrinsic scatter of $\sim$0.3 dex \citep{2007ApJ...660L..47N,2007ApJ...660L..43N}, which places strong constraints on the regulation of star formation processes of galaxies. The MS was originally defined for local star-forming galaxies, and later verified to be still present in higher redshifts but migrated dramatically to higher SFRs in higher redshift regimes. It implies that star-forming galaxies form their stars at a roughly steady state associated with stellar mass during most of their lifetime, and deviations
in star formation are short-lived until the galaxy is finally quenched and drops off from the MS \citep{2023ApJ...942...49C}. However, starburst galaxies are featured by intense star formation in a short period, so they are high above the MS shape. Passive galaxies with low star formation fall off the MS shape.    

Low Surface Brightness Galaxies (LSBGs) are traditionally defined as those galaxies with central surface brightness ($\mu_{0,B}$) at least one magnitude fainter than the night sky brightness ($\mu_{0,B} >$ 22.5 mag arcsec$^{-2}$) \citep{ 1994ApJ...426..135M,1995MNRAS.274..235D,doi:10.1146/annurev.astro.35.1.267,1997PASP..109..745B}. Galaxies satisfying this definition cover a wide range of physical properties, such as size, luminosity, morphology, color, and environment. They cover from dwarf irregulars to giant spirals, from blue to red, and live in both fields and galaxy clusters.
In contrast to galaxies with normal(high) surface brightnesses, the LSB disks tend to be poor in metal\citep{1994ApJ...426..135M} and dust, rich in neutral hydrogen gas (H{\sc{i}}) but have low star formation rates. They are a heterogeneous collection of galaxies from diverse origins, and the mechanisms of star formation and evolution of different LSBG populations are still in question. 
In order to explore the star formation of dwarf LSBGs, \citet{2017ApJ...851...22M} studied a sample of dwarf LSBGs in the SFR - M$_{*}$ plane, proposing that dwarf LSBGs are an entirely different population deviating from the MS galaxies. Differing from dwarf LSBGs, giant LSBGs (GLSBGs; \citet{1987AJ.....94...23B,1989ApJ...341...89I,1990ApJ...360..427B,1993ApJ...417..114S}) have extremely large, diffuse disks. They are very rare in the local universe (z $<$ 0.1) and represent the large-size end of the number density distribution of normal-size spiral galaxies \citep{2023MNRAS.520L..85S}. GLSBGs have fundamentally different origins, star formation histories and evolutionary paths from dwarf and other LSBGs. So far, few studies have been dedicated to the investigation of 
 a sample of GLSBGs in the MS plane. In this work, we would study the GLSBGs in the MS plane in order to explore the underpinning star formation processes of GLSBGs.
 
We collect GLSBGs and describe the multi-wavelength imaging data in Section ~\ref{subsec:data}, and estimate the star formation rates (SFRs) and stellar masses (M$_{*}$) in Section ~\ref{subsec:mstar-sfrs}. We study GLSBGs in the SFR - M$_{*}$ plane in Section ~\ref{subsec:ms_global}, H{\sc{i}} gas in Section ~\ref{subsec:HI} and constrain H$_{2}$ gas in Section ~\ref{subsec:H2}. The inner and outer parts of each GLSBG are separately studied in the SFR - M$_{*}$ plane, and the formation mechanism is discussed in Section ~\ref{sec:discus}. In Section ~\ref{sec:sum}, the whole work is summarised. All the distance-dependent quantities in this work are computed assuming  a $\Lambda$CDM cosmology with H$_{0}$ = 70 km s$^{-1}$ Mpc$^{-1}$, $\Omega_{m}$=0.3 and $\Omega_{\Lambda}$ = 0.7. The \citet{2003ApJ...586L.133C} initial mass function (Chabrier IMF) is assumed and AB magnitudes are used throughout the paper. 

\section{Data}
\subsection{Sample}\label{subsec:data}
It is estimated that there are around 12,700 GLSBGs in the entire sky out to z$<$0.1\citep{2023MNRAS.520L..85S}, but only a handful of them (less than 100)  have been truly observed and identified in the literature so far \citep{1985ApJS...59..115K,1993ApJ...417..114S,1994AJ....107..530M,1995AJ....109..558S,1995AJ....109.2019M,1997AJ....114.1858P,1998AJ....116.1650S,2001AA...365....1M,2007ApJ...663..908R,2015MNRAS.447.3649M,2015ApJ...815L..29G,2016ApJ...826..210H,2016AA...593A.126B,2017MNRAS.464.2741M}. We collected 27 designated GLSBGs (Table~\ref{tab:GLSBGs1}) observed by both GALEX ultraviolet (FUV, NUV) \citep{2007ApJS..173..682M,2014Ap&SS.354..103B,2014AdSpR..53..900B} and DECaLs \citep{2019AJ....157..168D} optical ($g$, $r$) imaging surveys from the literature as the sample.  Seventeen galaxies of the sample (denoted as S17; the first 17 galaxies in Table ~\ref{tab:GLSBGs1}) meet the `diffuseness index' criterion $\mu_{B}$(0) + 5log (Rs) $>$ 27.0 by \citet{1995AJ....109..558S} where $\mu_{B}$(0) is the disk B-band central surface brightness in mag arcsec$^{-2}$ and Rs is the disk scale length in $h^{-1}$ kpc based on a distance scale with $H_{0}$ = 100 km s$^{-1}$ Mpc$^{-1}$. The rest ten galaxies of the sample (denoted as S10; the last 10 galaxies in Table ~\ref{tab:GLSBGs1}) are from \citet{1998AJ....116.1650S} that defined GLSBGs with low central surface brightness ($\mu_{0}$(B) $>$ 22 mag arcsec$^{-2}$) or mean surface brightness ($\mu_{\rm eff}$(B) $>$ 24 mag arcsec$^{-2}$), large diameter (D $>$ 20 kpc) and high H{\sc{i}} mass (M$_{\rm HI}>$3.1$\times$ 10$^{9}$ M$_{\odot}$). Although the two selection criteria are seemingly different, they fundamentally aim to pick out galaxy populations that are large in size (stellar mass) and simultaneously have low surface brightness. However, the `diffuseness index' criterion only constrains a GLSBG in size and surface brightness, lacking some constraint for the stellar mass or luminosity, so some galaxies could meet the `diffuseness index' criterion by their large size and low surface brightness, but are not conventionally giant in luminosity or stellar mass, such as UGC 9024 in our sample. Although UGC 9024 is not a giant in stellar mass, it is indeed identified by the `diffuseness index' criterion and studied as a GLSBG  \citet{1994AJ....107..530M,1995AJ....109..558S,1995AJ....109.2019M,1997AJ....114.1858P,2001AA...365....1M,2007ApJ...663..908R}. Therefore, we also include UGC 9024 in our sample, but mark it (a blue filled diamonds) in figures. 

\begin{deluxetable*}{lcccccr}
\tablenum{1}
\tablecaption{The 27 GLSBGs collected from the literature \label{tab:GLSBGs1}}
\tablewidth{0pt}
\tablehead{
\colhead{GLSBG} & \colhead{RA} & \colhead{DEC} &
\colhead{velocity} & 
\colhead{MH{\sc{i}}} & \colhead{err$_{\rm MHI}$} &
\colhead{ref \textsuperscript{a}} \\
\colhead{Name} & \colhead{deg} & \colhead{deg} &
\colhead{(km/s)} & 
\colhead{10$^{10}\rm$ M$_{\odot}$} &
\colhead{10$^{10}\rm$ M$_{\odot}$} &
\colhead{} 
}
\decimalcolnumbers
\startdata
0052-0119 & 13.7872 & -1.0465 & 13197 & - & - &3\\
  0221+0001 & 36.0057 & 0.2523 & 37263 & - & - &3\\
  0237-0159 & 40.0461 & -1.7744 & 12701 & - & - &3 \\
  1034+0220 & 159.3653 & 2.0895 & 21335 & - & - &3\\
  1226+0105 & 187.3036 & 0.8178 & 23655 & 3.71 & -&9,3,14\\
  F533-3 & 334.3049 & 25.2130 & 12396 & 1.74 & - &10\\
  Malin-1 & 189.2473 & 14.3304 & 23793 & 4.57 & - &6,14,7,8\\
  Malin-2 & 159.9687 & 20.8470 & 13494 & 3.60 & 0.40 &14,1,2\\
  NGC4017 & 179.6901 & 27.4523 & 3428 & 1.70 & 0.35 &12,3,14\\
  NGC5533 & 214.0321 & 35.3437 & 3818 & 1.00 & 0.44 &12,3,14\\
  NGC5905 & 228.8472 & 55.5171 & 3350 & 2.45 & - &12,3,14\\
  NGC7589 & 349.5654 & 0.2612 & 8792 & 1.02 & - &10,14\\
  UGC1382 & 28.6711 & -0.1434 & 5767 & 1.70 & 0.10 &5\\
  UGC1752 & 34.0869 & 24.8883 & 17847 & 1.78 & - &4,10\\
  UGC6614 & 174.8118 & 17.1436 & 6353 & 2.50 & 0.20 &13,15,14,4,2,1\\
  UGC9024 & 211.6689 & 22.0700 & 2311.0 & 0.25 & - &13,15,3,14,10,11\\
  UM163\textsuperscript{b} & 352.6348 & -2.4625 & 10022.0 & 0.34 & 0.04 &2,1,3\\
  \hline
  UGC12740 & 355.4723 & 23.8150 & 10522 & 1.12 & 0.13 &4\\
  UGC12845 & 358.9245 & 31.8997 & 4880 & 0.58 & 0.25 &4\\
  UGC1455 & 29.6999 & 24.8925 & 5070 & 3.72 & 0.60 &4\\
  UGC1922 & 36.9410 & 28.2092 & 10894 & 3.20 & 0.40 &1,2,4\\
  UGC3968 & 115.6885 & 66.2584 & 6780 & 0.96 & - &4\\
  UGC4219 & 121.6783 & 39.0902 & 12433 & 2.95 & - &4\\
  UGC4422 & 126.9249 & 21.4791 & 4330 & 1.50 & 0.50 &1,2,4\\
  UGC4985 & 140.7210 & 21.9753 & 10180 & 3.02 & 0.35 &4\\
  UGC6754 & 176.6966 & 20.6754 & 7023.0 & 2.09 & 0.29 &4\\
  UGC905 & 20.4467 & 23.7809 & 11465.0 & 1.32 & 0.15 &4\\  
\enddata
\tablecomments{\\
a.  The ref No.s refer to the following papers.
1 - \citet{2017MNRAS.464.2741M},
2 - \citet{2015MNRAS.447.3649M},
3 - \citet{1995AJ....109..558S},
4 - \citet{1998AJ....116.1650S},
5 - \citet{2016ApJ...826..210H},
6 - \citet{1987AJ.....94...23B},
7 - \citet{2015ApJ...815L..29G},
8 - \citet{2016AA...593A.126B},
9 - \citet{1993ApJ...417..114S},
10 - \citet{2001AA...365....1M},
11 - \citet{2007ApJ...663..908R},
12 - \citet{1985ApJS...59..115K},
13 - \citet{1994AJ....107..530M},
14 - \citet{1997AJ....114.1858P}, and
15 - \citet{1995AJ....109.2019M}.\\
b. UM163 is also known as 2327-0244 in ref NOs. 3.
}
\end{deluxetable*}

\begin{deluxetable*}{lccccccccccc}
\tablenum{2}
\tablecaption{Measured properties of the 27 GLSBGs. \label{tab:GLSBGs2}}
\tablewidth{0pt}
\tablehead{
\colhead{GLSBG} & \colhead{FUV} & \colhead{err$_{\rm FUV}$} &
\colhead{NUV} & \colhead{err$_{\rm NUV}$} & 
\colhead{g}& \colhead{err$_{\rm g}$}  & \colhead{r} & \colhead{err$_{\rm r}$} &
\colhead{SFR\textsuperscript{a}} & 
\colhead{err$_{\rm SFR}$} &
\colhead{M$_{*}$\textsuperscript{a}}\\
\colhead{Name} &  \colhead{mag} & \colhead{mag} & 
\colhead{mag} & \colhead{mag} & 
\colhead{mag} & \colhead{mag}& \colhead{mag} & \colhead{mag} &
\colhead{M$_{\odot}$/$\rm yr$} & 
\colhead{M$_{\odot}$/$\rm yr$} &
\colhead{10$^{11}\rm$ M$_{\odot}$} 
}
\decimalcolnumbers
\startdata
0052-0119 & 20.690 & 0.102 & 18.826 & 0.042 & 14.467 & 0.003 & 13.676 & 0.002 & 0.06 & 0.010 & 2.02\\
0221+0001 & 18.914 & 1.303 & 18.559 & 0.057 & 16.485 & 0.002 & 15.755 & 0.002 & 2.84 & 3.553 & 3.54 \\
0237-0159 & 17.653 & 0.701 & 17.323 & 0.033 & 15.287 & 0.002 & 14.663 & 0.002 & 0.94 & 0.653 & 0.89\\
1034+0220 & 19.509 & 0.056 & 18.738 & 0.660 & 15.868 & 0.001 & 15.129 & 0.001 & 0.50 & 0.053 & 1.48 \\
1226+0105 & 18.412 & 0.968 & 18.046 & 0.047 & 15.731 & 0.002 & 15.087 & 0.002 & 1.57 & 1.479 & 1.40\\
F533-3 & 18.162 & 1.001 & 17.758 & 0.041 & 14.961 & 0.011 & 14.248 & 0.007 & 0.56 & 0.544 & 1.35\\
Malin-1 & 19.515 & 0.063 & 18.912 & 0.702 & 16.764 & 0.016 & 15.994 & 0.011 & 0.63 & 0.070 & 0.97 \\
Malin-2 & 17.402 & 0.614 & 17.062 & 0.020 & 14.285 & 0.004 & 13.582 & 0.002 & 1.34 & 0.824 & 1.98\\
NGC4017 & 14.952 & 0.199 & 14.646 & 0.008 & 12.833 & 0.001 & 12.383 & 0.001 & 0.79 & 0.177 & 0.27\\
NGC5533 & 15.930 & 0.301 & 15.582 & 0.143 & 12.142 & 0.002 & 11.389 & 0.001 & 0.40 & 0.128 & 1.08\\
NGC5905 & 14.789 & 0.178 & 14.534 & 0.008 & 12.145 & 0.001 & 11.490 & 0.001 & 0.87 & 0.179 & 0.57\\
NGC7589 & 17.458 & 0.030 & 17.112 & 0.156 & 14.354 & 0.004 & 13.679 & 0.003 & 0.53 & 0.040 & 0.88\\
UGC1382 & 17.709 & 0.722 & 17.439 & 0.025 & 13.714 & 0.002 & 12.919 & 0.001 & 0.18 & 0.127 & 0.79\\
UGC1752 & 17.578 & 0.799 & 17.431 & 0.398 & 14.946 & 0.004 & 14.261 & 0.003 & 1.91 & 1.502 & 1.80\\
UGC6614 & 16.535 & 0.435 & 15.991 & 0.013 & 13.718 & 0.002 & 13.067 & 0.001 & 0.63 & 0.281 & 0.93\\
UGC9024\textsuperscript{b} & 16.599 & 0.431 & 16.387 & 0.021 & 14.983 & 0.003 & 14.599 & 0.002 & 0.08 & 0.035 & 0.010\\
UM163\textsuperscript{c} & 16.964 & 0.055 & 16.449 & 0.016 & 13.402 & 0.003 & 12.802 & 0.002 & 1.09 & 0.106 & 1.81\\
\hline
UGC12740 & 17.710 & 0.796 & 17.284 & 0.414 & 15.235 & 0.006 & 14.717 & 0.004 & 0.58 & 0.457 & 0.31\\
UGC12845 & 16.388 & 0.423 & 16.164 & 0.02 & 13.509 & 0.003 & 12.895 & 0.002 & 0.42 & 0.182 & 0.45\\
UGC1455 & 15.794 & 0.377 & 15.735 & 0.195 & 12.619 & 0.015 & 11.858 & 0.010 & 0.80 & 0.313 & 1.56\\
UGC1922 & 17.220 & 0.046 & 16.537 & 0.036 & 13.509 & 0.011 & 12.955 & 0.007 & 0.98 & 0.088 & 2.04\\
UGC3968 & 16.670 & 0.458 & 16.368 & 0.232 & 13.955 & 0.003 & 13.293 & 0.002 & 0.64 & 0.298 & 0.47\\
UGC4219 & 16.807 & 0.512 & 16.504 & 0.023 & 14.132 & 0.005 & 13.488 & 0.003 & 1.89 & 0.977 & 1.10\\
UGC4422 & 15.332 & 0.249 & 14.970 & 0.013 & 12.444 & 0.001 & 11.849 & 0.001 & 0.87 & 0.237 & 0.92\\
UGC4985 & 16.983 & 0.524 & 16.440 & 0.019 & 13.617 & 0.003 & 12.922 & 0.002 & 1.06 & 0.564 & 2.22\\
UGC6754 & 15.844 & 0.303 & 15.496 & 0.012 & 12.807 & 0.003 & 12.139 & 0.002 & 1.45 & 0.467 & 1.75\\
UGC905 & 19.138 & 1.585 & 18.445 & 0.618 & 14.788 & 0.005 & 13.997 & 0.004 & 0.19 & 0.282 & 1.17\\
\enddata
\tablecomments{\\
a. SFRs and M$_{*}$ are in the Chabier IMF. SFRs are from FUV flux with no correction for dust attenuation. \\
b. This galaxy is identified as a giant LSBG according to the `diffuseness index' criterion in references of No.s 3, 10, 11,13, 14, and 15 in Table ~\ref{tab:GLSBGs1}, although it does not seem to be a giant in stellar mass in this table.\\
c. UM163 is also known as 2327-0244 in \citet{1995AJ....109..558S}.
}
\end{deluxetable*}

\begin{figure}[htbp]
\centering
\includegraphics[width=0.8\textwidth]{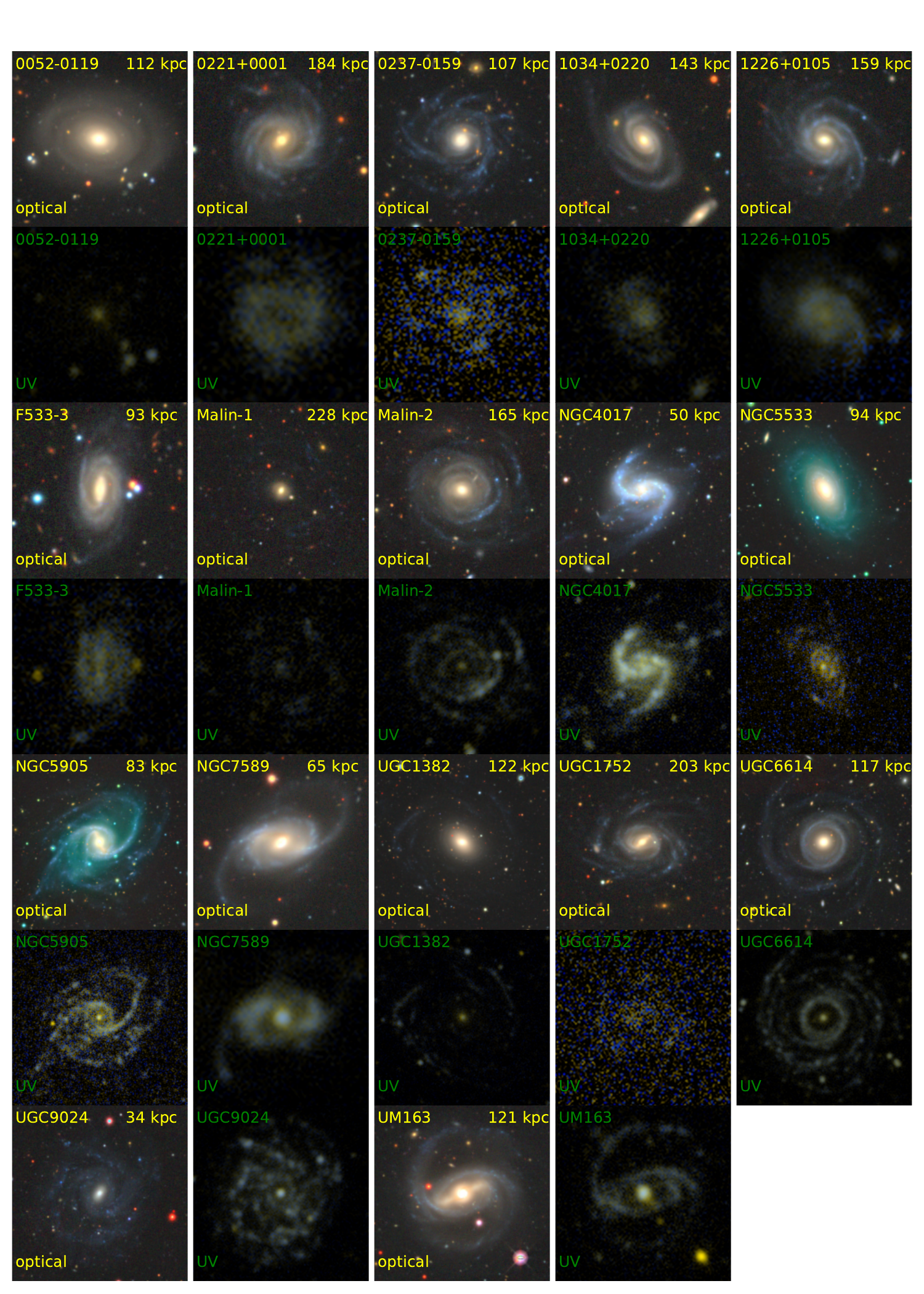}
\caption{Images of the 17 GLSBGs definitely meeting the `diffuseness index' criterion. For each galaxy, we show its optical and UV images, respectively, in individual frames of the same physical side size labeled at the top right of the optical frame. The optical image is the $g$-,$r$-, and $z$-composite image from DECaLs, except for NGC5533 and NGC5905 which have no $z$-band images from DECaLs, and the UV image is the FUV- and NUV-composite image from GALEX. \label{fig:glsbg_set1}}
\end{figure}

\begin{figure}[htbp]
\centering
\includegraphics[width=0.8\textwidth]{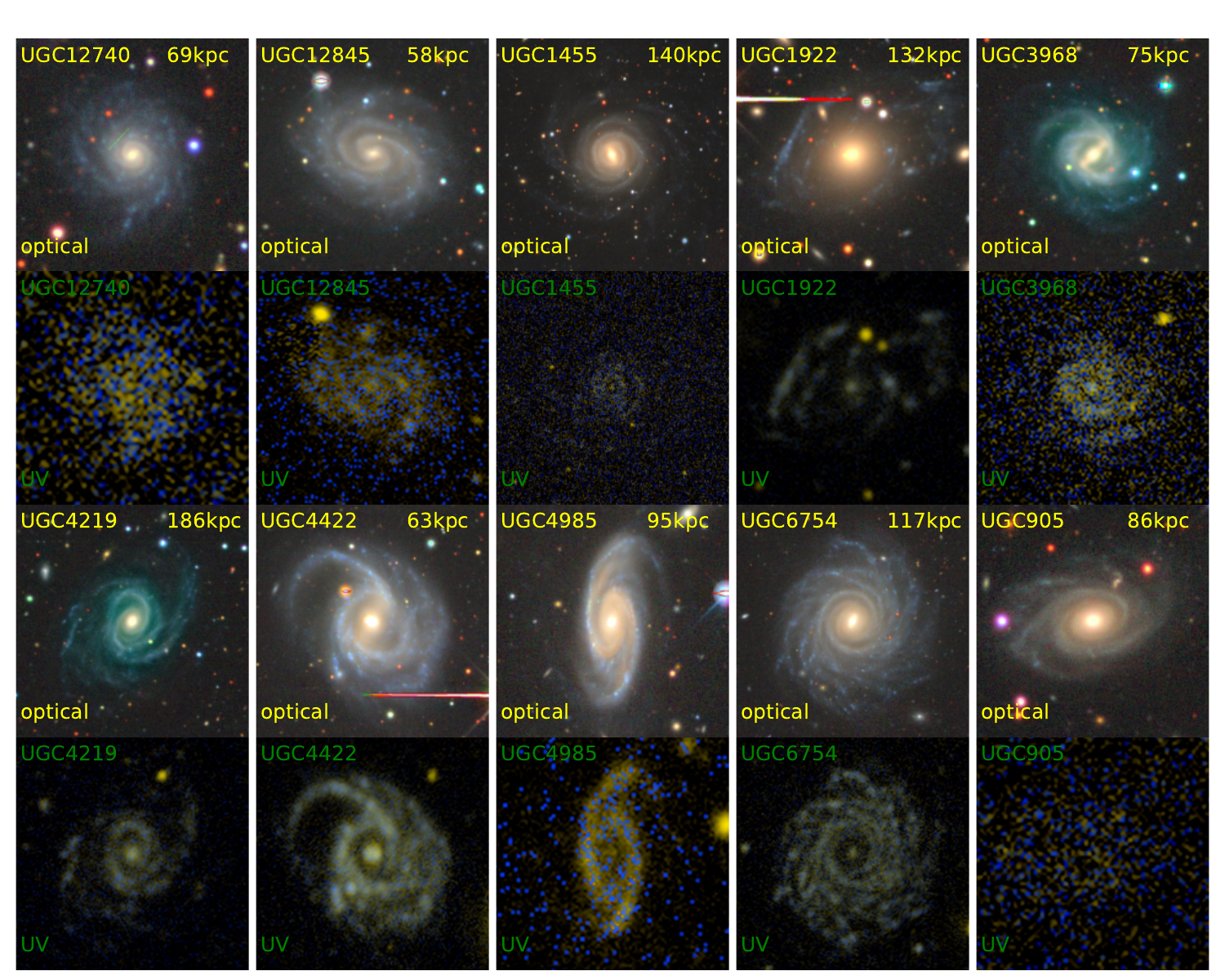}
\caption{Images of the rest 10 GLSBGs from \citet{1998AJ....116.1650S}. The illustration for this figure is the same as Figure ~\ref{fig:glsbg_set1}. UGC3968 and UGC4219 have no $z$-band images from DECaLs so their optical images are the composite of only $g$- and $r$-band images from DECaLs. \label{fig:glsbg_set2}}
\end{figure}

\subsection{SFR and Stellar Mass} \label{subsec:mstar-sfrs}
We obtain image frames of FUV, NUV (GALEX DR6), $g$, $r$, and $z$ bands (DECaLs DR9) for each GLSBG in our sample. Then, we match the Point Spread Functions (PSFs) of the image frames of FUV, $g$, $r$, and $z$ bands to the PSF of NUV band that is the lowest among the five bands. The surrounding bright stars of the target GLSBG are masked from the PSF-matched image frames to reduce light pollution. The reduced images of the five bands for each GLSBG are simultaneously fed into the SExtractor code to measure the flux in each band. In the dual mode of SExtractor, the extraction and the Kron aperture for photometry of the target GLSBG are first defined in the reduced $r$-band image frame, and then applied in image frames of the other four bands. The measured fluxes are next converted to magnitudes. All the magnitudes are corrected for the effects of Galactic extinction via A$_{\lambda}$ = R$_{\lambda}$E(B-V) where R$_{\lambda}$ is the ratio of the total absorption A$_{\lambda}$ to the reddening E(B-V) along the line of sight to an object. For each GLSBG, the E(B-V) value are estimated by \citet{2011ApJ...737..103S}(SF11). In the optical $g$, $r$, and $z$ bands, the extinction coefficients R$_{\lambda}$ ($\lambda = g, r, z$) are, respectively, provided in SF11 at R$_{\rm V}$=3.1.In FUV and NUV bands, the extinction coefficients of R$_{\rm FUV}$ = 8.29 and R$_{\rm NUV}$ = 8.87 are assumed. 

Star formation rate (SFR) can be estimated from the FUV flux for each GLSBG. As the extended LSB disks are acknowledged to be deficient in dust \citet{2007ApJ...663..908R},  at first we do not correct the effect of internal dust extinction for FUV flux. Adopting the luminosity distance ($d_{L}$) calculated at the galaxy's redshift (z; corresponding to the radial velocity in Table ~\ref{tab:GLSBGs1}) in the $\Lambda$CDM cosmology, FUV flux is converted to FUV luminosity (L$_{\rm FUV}$) that is subsequently converted to SFR via the empirical relation between L$_{\rm FUV}$ and SFR \citep{2007ApJS..173..267S}. In addition, we still derive a set of dust-corrected SFRs for GLSBGs in our sample for comparison. We calculate FUV attenuation (A$_{\rm FUV}$) from NUV - FUV color by using the prescription in \citet{2007ApJS..173..267S}. Then, the dust-corrected FUV fluxes are used to estimate dust-corrected SFRs. It is worth noting that during dust attenuation correction for FUV fluxes, we find that the estimated A$_{\rm FUV}$ have extremely large errors (even larger than the values themselves) that further introduce very large uncertainties to the dust-corrected SFRs that have errors at least 2$\sim$3 times larger than the SFR values themselves (see the error bars in the right panel of Figure ~\ref{fig:sfms}). Therefore, we prefer the SFRs without dust correction (Table ~\ref{tab:GLSBGs2}; the left panel of Figure ~\ref{fig:sfms}).  
 
Stellar mass (M$_{*}$) of each GLSBG is derived from fitting the UV-optical SEDs by using CIGALE \citep{2009A&A...507.1793N,2019A&A...622A.103B}. We opt to use \citet{2003MNRAS.344.1000B} (BC03) stellar population model, the Chabrier IMF, the two-component exponential star formation histories, and the modified dust attenuation laws \citep{2019A&A...622A.103B}. For the parameters of the adopted two-component exponential model, the formation time of the old population spans a range from 10 Gyr to 12 Gyr before the observation epoch, with e-folding times that span a range from 500 Myr to 5 Gyr. For the young population, the formation time spans from 500 Myr to 5 Gyr with a constant e-folding time of 20 Gyr. Thus, the star formation histories adopted look like old exponentials with distinct decay times superimposed by a relatively flat burst.
The derived M$_{*}$ are at 2.68$\times$10$^{10}$ -3.54$\times$ 10$^{11}$ M$_{\odot}$ for the sample except for the galaxy UGC 9024 which has M$_{*} \sim$ 9.92 $\times$ 10$^{8}$ M$_{\odot}$, two orders of magnitude lower than the average M$_{*}$ of the other GLSBGs (Table ~\ref{tab:GLSBGs2}). It is not a conventional giant in stellar mass. However, as we mentioned in Section ~\ref{subsec:data}, this galaxy is identified as a GLSBG according to the `diffuseness index' criterion in several major papers. 

\section{Results} 
\subsection{GLSBGs in the SFR-stellar mass plane}\label{subsec:ms_global}
\begin{figure}[htbp]
\includegraphics[width=1.0\textwidth]{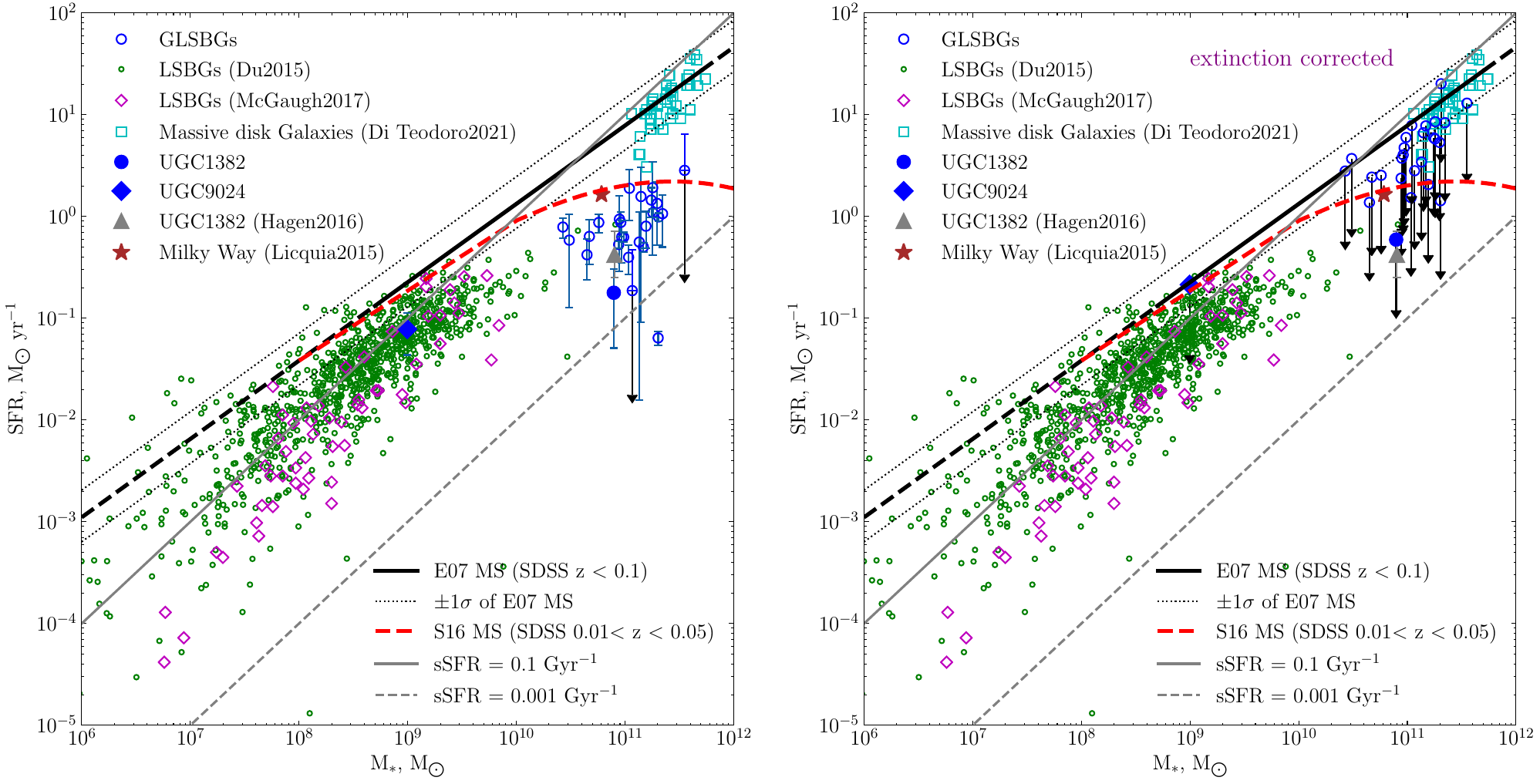}
\caption{Star formation rate as a function of stellar mass. In both panels, GLSBGs are represented by the big blue open circles with error bars. For comparisons, the H{\sc{i}}-rich LSBGs from \citet{2015AJ....149..199D} (small green circles), the dwarf LSBGs from \citet{2017ApJ...851...22M} (magenta open diamonds), massive spiral galaxies from \citet{2021MNRAS.507.5820D} (cyan open squares), the Milky Way with M$_{*}$ = 6.08 $\times$ 10$^{10}$ M$_{\odot}$ and SFR = 1.65 M$_{\odot}$ yr$^{-1}$ from \citet{2015ApJ...806...96L} (red filled star), and the GLSBG UGC1382 from \citet{2016ApJ...826..210H} (the grey filled triangle with error bar) are displayed as well. For a direct comparison, the UGC 1382 with our measurements in this work is highlighted by the filled blue circle. The black solid line shows the main sequence (MS) trajectory defined for the local SDSS star-forming galaxies (0.04 $<$ z $<$ 0.1, 5$\times$10$^{8} <$ M$_{*} <$ 5$\times$10$^{11}$M$_{\odot}$ ) by \citet{2007A&A...468...33E}, with the extending black, dashed lines representing its linear extrapolation to both lower mass and higher mass ends (E07 MS), and the two black dotted lines show the 68\% confidence level of the E07 MS. The red dashed curve is the star formation MS trajectory defined for local SDSS galaxies (0.01 $<$ z $<$ 0.05, 10$^{8} <$ M$_{*}<$10$^{12}$M$_{\odot}$) by \citet{2016MNRAS.462.1749S} (S16 MS). The grey solid and dashed lines represent a constant sSFR value of 0.1 and 0.001 Gyr$^{-1}$. The length of the black downward arrow on some point does not refer to the actual lower error of the point, and it only means that the original lower error bar below the data point is so large that it stretches far below the figure boundary. In the left panel, SFRs are estimated from FUV flux without correction for the internal dust extinction, while in the right panel, SFRs are from dust-corrected FUV flux however, the correction is quite uncertain for this population of low surface brightness galaxies which should be very low in dust so it definitely introduces very large errors. The Chabrier IMF is assumed. \label{fig:sfms}}
\end{figure}

 We show the GLSBGs (blue open circles) in the SFR - M$_{*}$ plane (Figure ~\ref{fig:sfms}). The SFRs of GLSBGs in the left panel are without dust correction and those in the right panel are dust-corrected (Section ~\ref{subsec:mstar-sfrs}). 
 
 In both panels, GLSBGs are obviously below the MS defined for the local SDSS star-forming galaxies (SFGs; 0.04 $<$ z $<$ 0.1, 5$\times$10$^{8} <$ M$_{*} <$ 5$\times$10$^{11}$ M$_{\odot}$ ) by \citet{2007A&A...468...33E} (black solid + dashed line; E07 MS). We also show the MS from \citet{2016MNRAS.462.1749S} that fitted SFRs as a function of M$_{*}$ with a third-order polynomial (red dashed curve; S16 MS) for the local SDSS SFGs (0.01 $<$ z $<$0.05, 10$^{8} <$ M$_{*} <$ 10$^{12}$ M$_{\odot}$). The S16 MS traces a flattening relation at the high mass regime (M$_{*} >$ 10$^{10}$ M$_{\odot}$), where the S16 MS is lower than E07 MS in SFR. For example, it is 0.16, 0.59, and 1.39 dex lower in SFR than E07 MS at M$_{*}\sim$ 10$^{10}$, 10$^{11}$, and 10$^{12}$ M$_{\odot}$ , respectively. 
 
 In the left panel, the majority of GLSBGs are clearly below the S16 MS curve while the GLSBGs move up to straddle the S16 MS curve in the right panel due to dust extinction correction. However,  as we mentioned in Section ~\ref{subsec:mstar-sfrs}, the dust extinction correction for these GLSBGs that are supposed to be poor in dust is so uncertain that the frustrating error bars of dust-corrected SFRs stretch downward the full figure frame and even cross the lower boundary of the figure frame in the right panel. Therefore, the positions of the blue open circles in the right panel represent the upper limits of the dust-corrected GLSBGs, and the length of the black downward arrow on each blue open circle does not represent any meaningful value, but only means that the lower error bar is too long to be fully displayed in the current frame of the right panel.  Due to the very large uncertainties of the dust-corrected SFRs, we prefer the conclusions deduced from the left panel (no dust correction).

  The GLSBG UGC1382 measured by  \citet{2016ApJ...826..210H} (the grey filled triangle; UGC1382-H) is shown to compare with our measurement for UGC 1382 (the blue filled circle). It shows our measurement is consistent with the measurement by  \citet{2016ApJ...826..210H} within the errors. For comparison with giant galaxies with normal(high) surface brightness, we show the massive, disk-dominated galaxies from \citet{2021MNRAS.507.5820D} (cyan open squares). The normal massive galaxies are clearly on the trajectory of E07 MS, and above the S16 MS at the high mass regime. In contrast, the GLSBGs do not behave similarly to MS galaxies and the normal massive galaxies in both SFR - M$_{*}$ panels because of much lower specific SFR (sSFR = SFR/M$_{*}$). The sSFRs of the GLSBGs are below 0.1 Gyr$^{-1}$ (grey solid line) which is the characteristic sSFR value for SFGs of M$_{*}$ = 10$^{10}$M$_{\odot}$ at z = 0, and above the constant line representing sSFR = 0.001 Gyr$^{-1}$ (grey dashed line). 
Despite normal massive galaxies, we also display the H{\sc{i}}-rich dwarf LSBGs from \citet{2015AJ....149..199D,2019MNRAS.483.1754D} (small green circles) and the dwarf LSBGs from \citet{2017ApJ...851...22M} (magenta open diamonds). The dwarf LSBGs are slightly below the star formation MS, particularly those at the low mass regime (M$_{*} <$ 10$^{8}$ M$_{\odot}$). They have higher sSFR than GLSBGs.
  
  Previous studies have shown that distances from MS relate to molecular gas contents and star formation efficiencies (SFE) of galaxies. Galaxies above the MS (mainly the starbursting galaxies) have high sSFRs because of the enhanced molecular gas mass fractions and increased SFEs, but galaxies below the MS (mainly the bulge-dominated galaxies) have low sSFRs and are observed to be low in both molecular gas mass fractions and SFEs \citep{2011MNRAS.415...61S,2012ApJ...758...73S}.
The GLSBGs in this work have low molecular gas content (discussed in \ref{subsec:H2}) and SFEs, which lead to their low sSFRs.

\subsection{Atomic gas versus stellar mass}\label{subsec:HI}
\begin{figure}[htbp]
\centering
\includegraphics[width=0.6\textwidth]{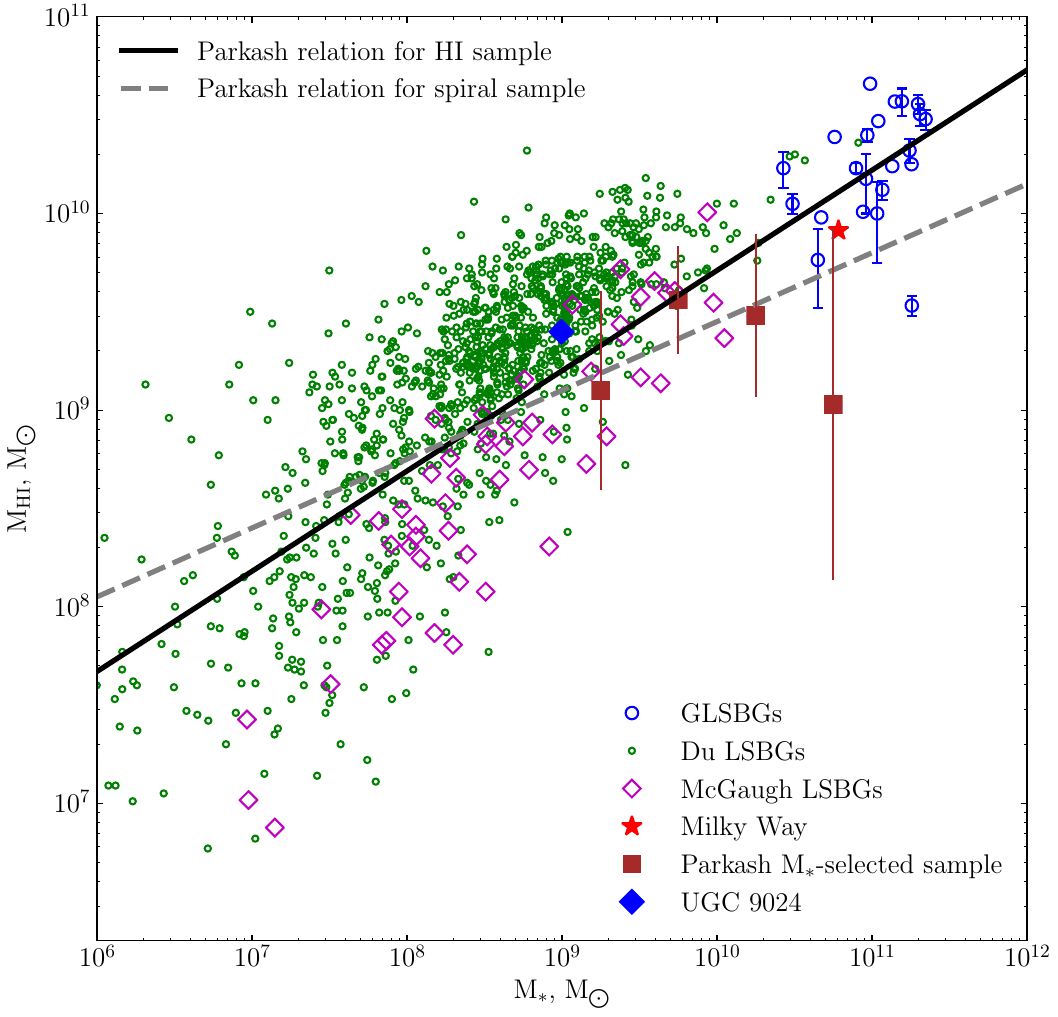}
\caption{Relation between H{\sc{i}} mass and stellar mass of the GLSBGs (blue open circles). For comparisons, the H{\sc{i}}-selected LSBGs from \citet{2015AJ....149..199D} (green crosses), LSBGs from \citet{2017ApJ...851...22M} (magenta open diamonds), and the M$_{*}$-selected sample of galaxies binned with a width of log M$_{*}$ = 0.5 (brown filled squares with error bars) from \citet{2018ApJ...864...40P}, and the Milky Way (red filled star) with M$_{\rm HI}$ = 8.2$\times$ 10$^{9}$ M$_{\odot}$ from \citep{2006A&A...459..113M} are displayed as well. The black solid line represents the relation followed by the H{\sc{i}}-selected sample of galaxies with 1$\sigma$ scatter of 0.5 dex, and the grey dashed line shows the relation followed by the spiral galaxies out of the M$_{*}$-selected sample with 1$\sigma$ scatter of 0.4 dex from \citet{2018ApJ...864...40P}. 
\label{fig:HI-mstar}}
\end{figure}

We collected H{\sc{i}} masses (M$_{\rm HI}$) for 23 out of these 27 GLSBGs from the literature \citep{1995AJ....109..558S,1997AJ....114.1858P,1999AJ....118..765P,1998AJ....116.1650S,2016ApJ...826..210H,2017MNRAS.464.2741M}. The M$_{\rm HI}$ values are tabulated in Table ~\ref{tab:GLSBGs1}. These GLSBGs are all H{\sc{i}}-rich systems with M$_{\rm HI}>$ 10$^{10}$ M$_{\odot}$ (0.955 - 4.57 $\times$ 10$^{10}$ M$_{\odot}$) for 20 out of 23 galaxies and M$_{\rm HI}\sim$ 2.5 - 5.8 $\times$ 10$^{9}$ M$_{\odot}$ for the rest 3 galaxies.   
We show these GLSBGs (blue open circles) in the M$_{\rm HI}$ - M$_{*}$ plane (Figure ~\ref{fig:HI-mstar}). GLSBGs are apparent to follow the trend (black solid line) defined by \citet{2018ApJ...864...40P} for a sample of H{\sc{i}}-rich galaxies, but are clearly beyond the trajectory (grey solid line) defined by \citet{2018ApJ...864...40P} for M$_{*}$-selected galaxies (brown filled squares). It implies that these GLSBGs have high H{\sc{i}} gas mass fractions (f$_{\rm HI}$ = M$_{\rm HI}$/M$_{*}$). Out of the 23 GLSBGs, 21 have f$_{\rm HI} >$0.1. In Figure ~\ref{fig:HI-mstar-ch}, these GLSBGs (blue open circles) show higher f$_{\rm HI}$ than the S16 MS galaxies (blue filled circles) along M$_{*}$. 

Generally, galaxies with copious amounts of H{\sc{i}} gas should have the potential to form stars actively as H{\sc{i}} gas is the fundamental ingredient in the formation of stars. However, GLSBGs with high H{\sc{i}} gas fractions in this figure have very inactive star formation, indicating that the atomic hydrogen gas (H{\sc{i}}) has not been efficiently converted into the molecular hydrogen gas (H$_{2}$) which are the direct fuel for star formation. We will discuss the low efficiency of H{\sc{i}} -to- H$_{2}$ conversion in GLSBGs in Section ~\ref{subsec:H2}.  

\subsection{Constraints on Molecular hydrogen gas}\label{subsec:H2}
Molecular gas is seldom detected in the morjority of GLSBGs. For instance, CO emission is never detected in Malin 1, an iconic GLSBG and also the largest spiral galaxy, even though very deep and high sensitivity observations are performed in the recent study\citep{2022ApJ...940L..37G}.  Instead, only an upper limit (3$\sigma$) of H$_{2}$ mass (M$_{\rm H_{2}} <$ 7.4 $\times$ 10$^{9}$ M$_{\odot}$) is estimated for Malin 1. Aligning the current upper limit of H$_{2}$ mass and the observed H{\sc{i}} mass (Table ~\ref{tab:GLSBGs1}), we could constrain the upper limit of molecular-to-atomic gas mass ratio (R$_{\rm mol}$ = M$_{\rm H2}$ /M$_{\rm HI}$) for Malin 1 is 0.12. 
However, Molecular gas is indeed detected in a few GLSBGs via CO observations \citep{2005AJ....129.1849M,2006ApJ...651..853D,2010A&A...523A..63D}. For instance, CO line emission is detected from Malin 2, a typical GLSBG, and then converted to molecular gas mass of M$_{\rm H_{2}}$ = 4.9 - 8.3$\times$10$^{8}$ M$_{\odot}$ through the assumed CO-to-H$_{2}$ conversion factor \citep{2010A&A...523A..63D}. With its M$_{\rm H_{I}}$ = 3.6$\pm$0.4 $\times$ 10$^{10}$ M$_{\odot}$ \citep{1997AJ....114.1858P,2010A&A...523A..63D}, Malin 2 has a molecular-to-atomic gas mass ratio (R$_{\rm mol}$ = M$_{\rm H2}$ /M$_{\rm HI}$) of $\sim$ 0.01 - 0.02 which is relatively high for an LSB galaxy. 
In other few GLSBGs where CO emission is also detected, the molecular gas mass converted via the CO-to-H$_{2}$ conversion factor is in the range of 10$^{8}$ - 10$^{9}$ M$_{\odot}$ \citep{2010A&A...523A..63D}. As H{\sc{i}} gas mass is typically in the level of M$_{\rm H_{I}} \sim $10$^{10}$ M$_{\odot}$ for most GLSBGs (Table ~\ref{tab:GLSBGs1}), we can deduce the typical R$_{\rm mol}$ values for most GLSBGs should be $\sim$ 0.01 - 0.1, which is a sufficient range to include the cases of Malin 2 (R$_{\rm mol}\sim$ 0.01 - 0.1) and Malin 1 (R$_{\rm mol} <$ 0.12). In contrast, the R$_{\rm mol}$ for normal massive galaxies with comparable stellar masses (M$_{*} \sim$ 10$^{10}$ - 10$^{11.5}$ M$_{\odot}$) is $\sim$ 0.3 on average \citet{2011MNRAS.415...32S}, which is higher than GLSBGs.  

\begin{figure}[htbp]
\centering
\includegraphics[width=1.0\textwidth]{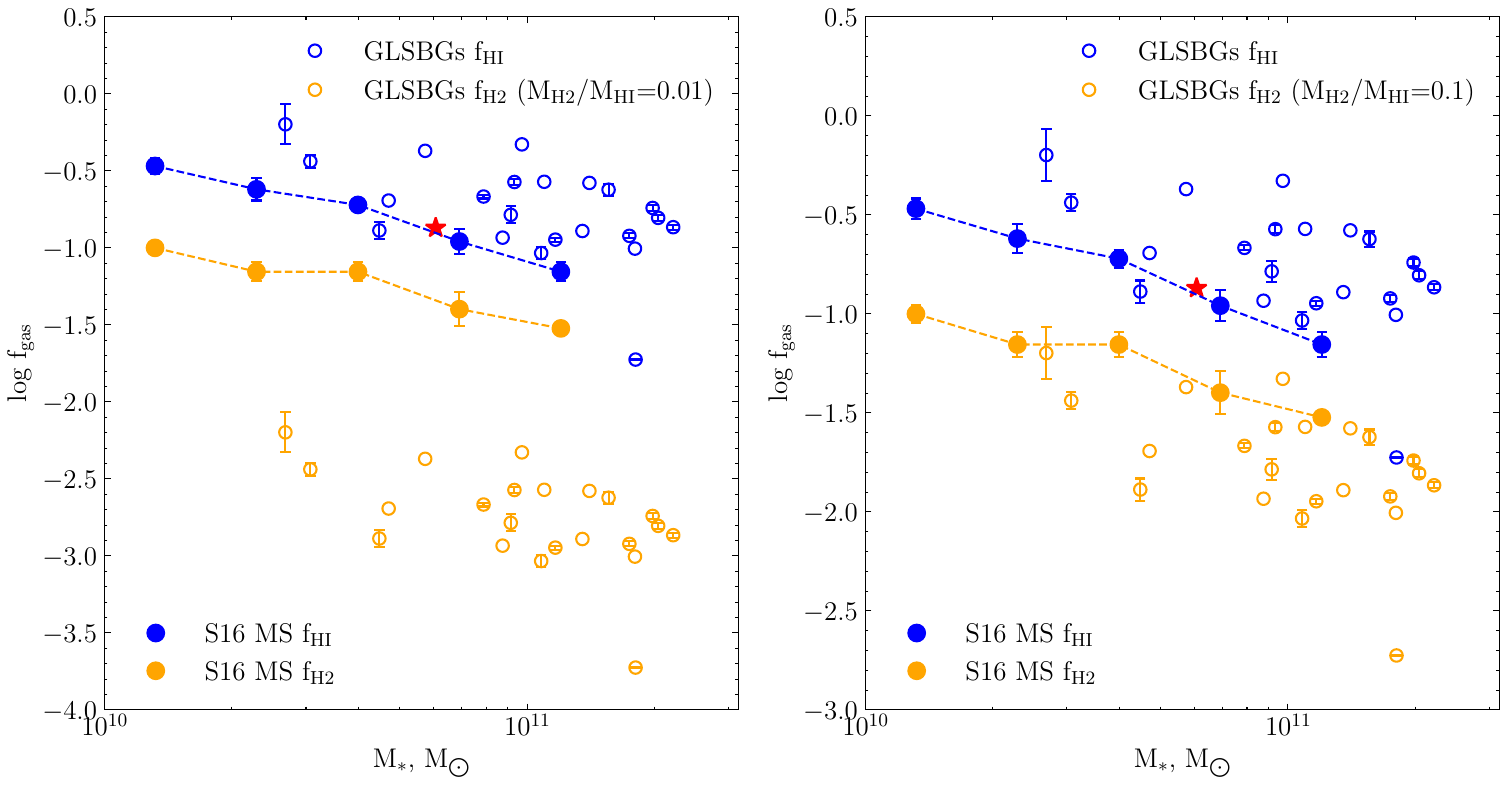}
\caption{Gas mass fraction versus stellar mass. The open circles represent GLSBGs and the filled circles represent the mean values of five bins in M$_{*}$ of S16 MS galaxies \citep{2016MNRAS.462.1749S}. The blue and orange colors represent H{\sc{i}} gas mass fractions (f$_{\rm HI}$ = M$_{\rm HI}$/M$_{*}$) and H$_{\rm 2}$ gas mass fractions (f$_{\rm H2}$ = M$_{\rm H2}$/M$_{*}$). The molecular-to-atomic ratio (R$_{\rm mol}$ = M$_{\rm H2}$ /M$_{\rm HI}$) for GLSBGs is assumed as 0.01 in the left panel and 0.1 in the right panel. The red filled star represents the Milky Way (M$_{\rm H_{2}}$ = 1.3$\times$ 10$^{9}$ M$_{\odot}$ and M$_{\rm HI}$ = 8.2$\times$ 10$^{9}$ M$_{\odot}$) from \citet{2006A&A...459..113M}. 
\label{fig:HI-mstar-ch}}
\end{figure}

According to the derived R$_{\rm mol}\sim$ 0.01 - 0.1 for GLSBGs, we calculate two sets of M$_{\rm H_{2}}$ values for GLSBGs at R$_{\rm mol}$ = 0.01 and 0.1, respectively. The corresponding sets of H$_{2}$ gas mass fractions (f$_{\rm H_{2}}$ = M$_{\rm H_{2}}$/M$_{*}$) of GLSBGs (orange open circles) are displayed in Figure ~\ref{fig:HI-mstar-ch}. It is obvious that at R$_{\rm mol}\sim$ 0.01 - 0.1, GLSBGs (orange open circles) have lower f$_{\rm H_{2}}$ than S16 MS galaxies (orange filled circles) which have R$_{\rm mol}\sim$0.3 \citep{2016MNRAS.462.1749S}. This strengthens that GLSBGs are deficient in molecular gas - the direct fuel of star formation.   



Given that GLSBGs are rich in H{\sc{i}} gas (see Section ~\ref{subsec:HI}) but deficient in H$_{\rm 2}$,
it is plausible that the copious amounts of H{\sc{i}} gas has not been efficiently converted into H$_{\rm 2}$. What causes the low efficiency of H{\sc{i}}-to-H$_{\rm 2}$ transition? The conversion of H{\sc{i}}-to-H$_{\rm 2}$ is closely related to metallicity, the equilibrium turbulent gas pressure, and the gas surface density \citep{2014MNRAS.437.3072K}. First, the low metal and dust content in GLSBGs makes the cooling of H{\sc{i}} gas difficult and results in inefficient conversion to H$_{2}$. Then, some research proposes that a minimum H{\sc{i}} surface density of $\Sigma_{\rm H_{I}} \sim$ 6 - 8 M$_{\odot}$ pc$^{-2}$ is required to provide enough shielding for H$_{\rm 2}$ formation against photodissociation \citep{2012ApJ...748...75L}. Once the minimum $\Sigma_{\rm H_{I}}$ is reached to shield H$_{\rm 2}$ from photodissociation, all excess H{\sc{i}} is converted into H$_{\rm 2}$ and the H$_{\rm 2}$ abundance increases. However, the H{\sc{i}} gas surface densities have been measured from the H{\sc{i}} maps for some GLSBGs in literature, showing that the peak H{\sc{i}} gas surface densities of these GLSBGs are indeed below the minimum value required to prevent the formed H$_{2}$ from photodissociation. For example, 
the peak value of $\Sigma_{\rm H_{I}}$ is below 2 M$_{\odot}$ pc$^{-2}$ for UGC1382 \citep{2016ApJ...826..210H}, 5 M$_{\odot}$ pc$^{-2}$ for Malin-1 \citep{2010A&A...516A..11L}, 5.5 M$_{\odot}$ pc$^{-2}$ for NGC7589 \citep{2010A&A...516A..11L}, 4 M$_{\odot}$ pc$^{-2}$ for UGC 6614 \citep{1997AJ....114.1858P},  which are far below the minimum required H{\sc{i}} surface density.  Although we do not have H{\sc{i}} radial surface density profiles for other individual GLSBGs yet,  GLSBGs are expected to be fundamentally consistent with each other in properties, so we deduce that the lower $\Sigma_{\rm H_{I}}$ for GLSBGs should result in low efficiency of H{\sc{i}}- H$_{\rm 2}$ transition and thus little supply of cold gas for star formation. However, Malin 2 is an exception. As we mentioned before, Malin 2 has a relatively high fraction of molecular gas among LSBGs,  and it is possibly attributed to either one or a combination of the following reasons including an underestimation of turbulent gas pressure, an incorrect CO-to-H$_{2}$ conversion factor, or a probable presence of some fraction of dark gas (invisible in CO and H{\sc{i}} lines) which could provide additional support for the observable H$_{2}$, shielding it from being photodissociated by UV radiation \citet{2014MNRAS.437.3072K}. 

\section{Discussion}\label{sec:discus}
 The star formation MS is conventionally defined based on star-forming disk galaxies, so the shape of MS should be dominantly regulated by the disks of galaxies. Recent studies proposed a turnover in MS shape at M$_{*} \sim$ 10$^{10}$ M$_{\odot}$, beyond which galaxies are no longer corresponding to higher SFRs due to the growth of bulge components which are quiescent and inactive in star formation activity, so they would drop off from the MS at the high mass regime. Since the disk form stars at a consistent level independent of bulge mass, the turnover should be mitigated if only the disk components are considered in high mass regime  \citep{2014ApJ...785L..36A,2015ApJ...808L..49G}. In terms of GLSBGs, they each host a significant bulge at the center of the extended diffuse disk (see Figure ~\ref{fig:glsbg_set1} and ~\ref{fig:glsbg_set2}), so the global deviations of them from MS (Figure ~\ref{fig:sfms}) are expected to be mainly from the bulge component. In order to validate this idea, we divide each GLSBG into inner and outer regions along the radius, and investigate the two regions separately in the SFR - M$_{*}$ plane in this section. 
  
In Section ~\ref{subsec:mstar-sfrs}, we have measured the half-light radius (R$_{\rm e,r}$) from the reduced (resolution-lowered to NUV and bright star-removed) $r$-band image for each GLSBG by using the SExtractor code. We adopt R$_{\rm e,r}$ as a demarcation to divide each GLSBG into inner (0 $<$ R $\leq$ R$_{\rm e}$) and outer (R$_{e} <$ R $\leq$ R$_{\rm Kron,r}$) regions on reduced FUV, NUV, $g$, $r$, and $z$-band images. R$_{\rm Kron,r}$ is the Kron aperture radius for flux measurement defined by SExtractor on the reduced $r$-band image. For each GLSBG, we measure annular fluxes of the inner and outer regions by using $Python\ Aperture\_photometry$ code, respectively, on the reduced FUV, NUV, $g$, $r$, and $z$-band images. Then we correct Galactic extinction for annular fluxes in each band, calculate the inner and outer SFRs from annular FUV fluxes of the inner and outer regions. For estimating the M$_{*}$ of the inner and outer regions of each GLSBG, we adopt the same method of SED-fitting as described in Section ~\ref{subsec:mstar-sfrs}. In terms of star formation history, for the inner region, we adopt a one-component exponental model of old population that has a formation time of 10 Gyr with the e-folding time spanning from 500 Myr to 5 Gyr. For the outer region, we adopt a one-component exponential model of young population that has a formation time spanning from 5 to 6 Gyr with the e-folding time from 800 Myr to 2 Gyr.

\begin{figure}[htbp]
\centering
\includegraphics[width=0.7\textwidth]{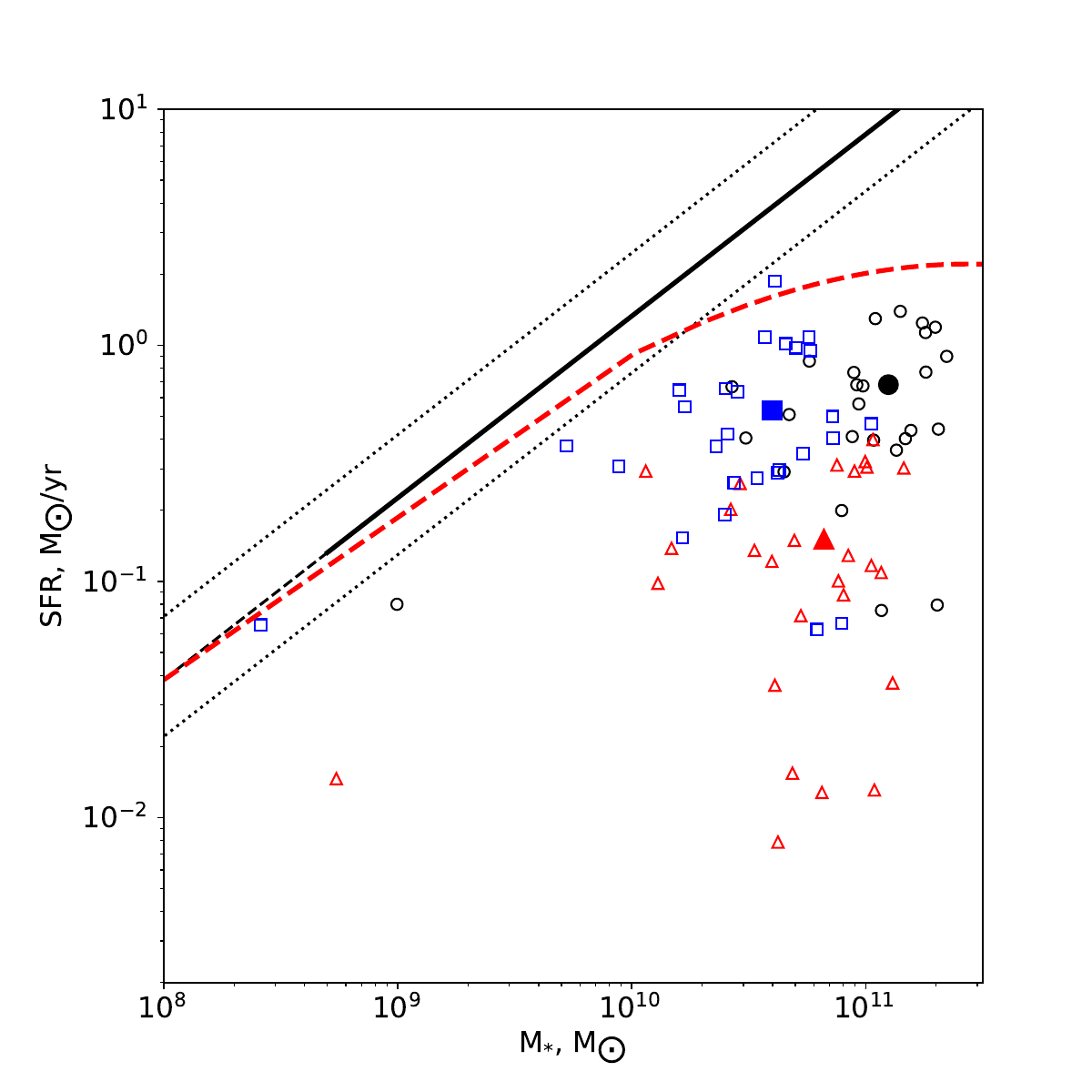}
\caption{Star formation rate (SFR) against stellar mass (M$_{*}$) for the inner and outer regions of the 27 GLSBGs. For each galaxy, the inner, the outer, and the entire galaxy are shown as an open red triangle, an open blue square, and an open black circle. The filled red triangle, blue square, and black circle, respectively, represent the mean values for the inner, outer, and entire galaxies. The black solid line represents the E07 MS \citep{2007A&A...468...33E}. The red dashed curve represents the S16 MS \citep{2016MNRAS.462.1749S}. The Chabier IMF is assumed. Note that the three isolating markers in the bottom left region in this figure are the inner, outer, and entire of UGC 9024, which is not a traditional giant galaxy in terms of stellar mass, but is identified as a GLSBG by the `diffuseness index' criterion from \citet{1995AJ....109..558S}. \label{fig:sfms_annuli_galaxyAll}}
\end{figure}

\begin{figure}[htbp]
\centering
\includegraphics[width=1.0\textwidth]{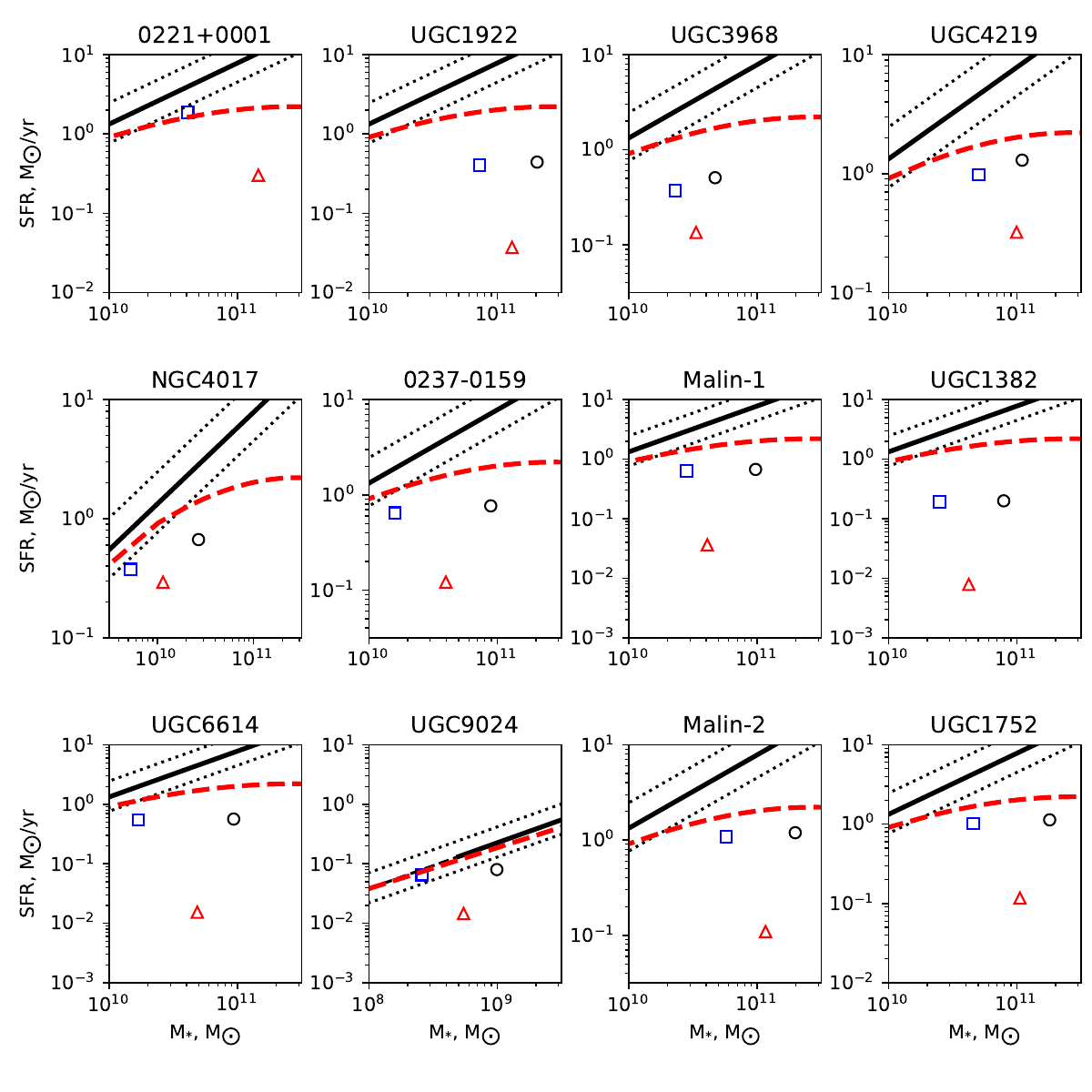}
\caption{Star formation rate (SFR) against stellar mass (M$_{*}$) for the inner (red triangle) and outer (blue square) parts of each GLSBG. The black circle represents the entire GLSBG. The black solid line represents the E07 MS \citep{2007A&A...468...33E}. The red dashed curve represents the S16 MS \citep{2016MNRAS.462.1749S}. The Chabier IMF is assumed.  \label{fig:sfms_annuli_galaxy12}}
\end{figure}

\begin{figure}[htbp]
\centering
\includegraphics[width=1.0\textwidth]{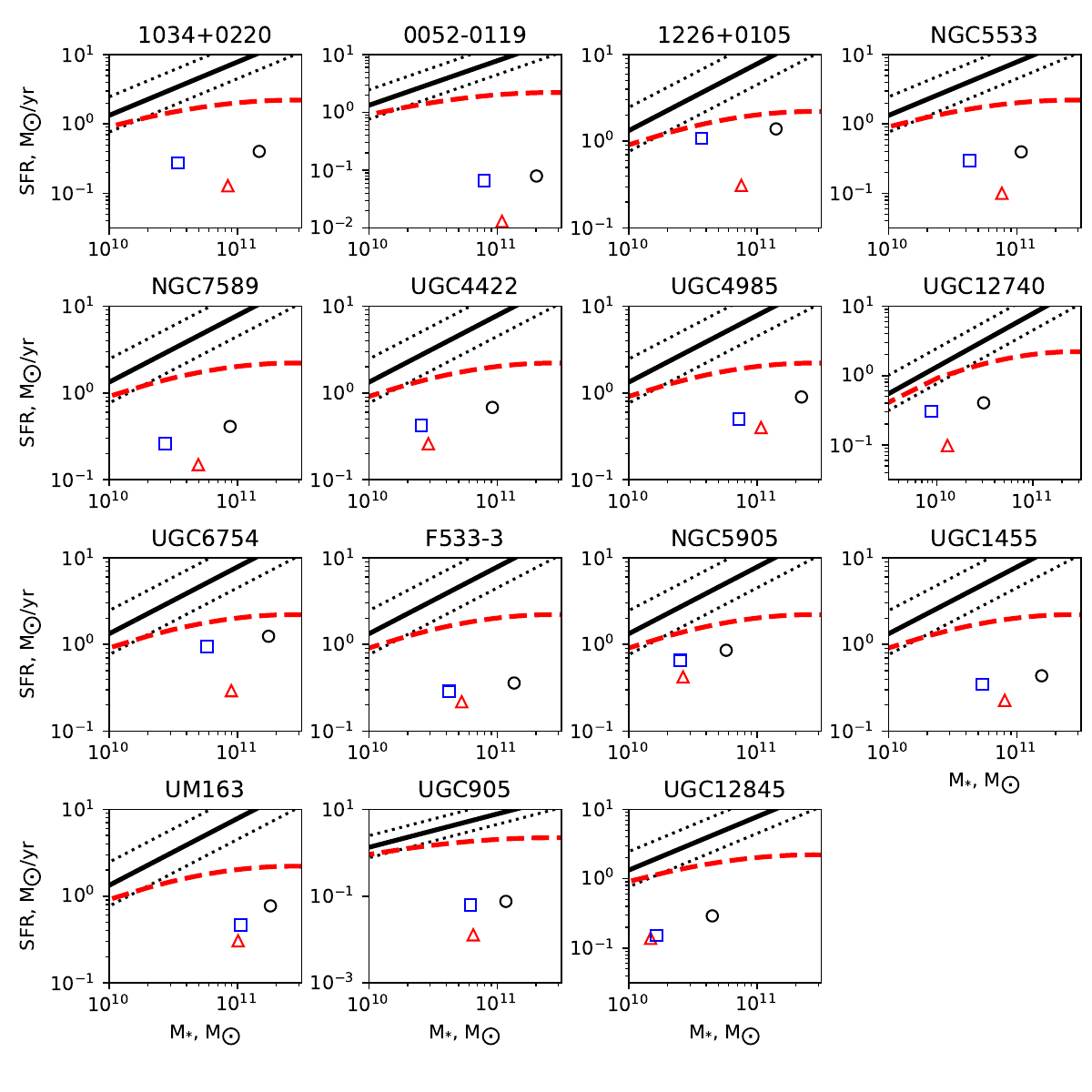}
\caption{Continued to Figure ~\ref{fig:sfms_annuli_galaxy12}.  \label{fig:sfms_annuli_galaxy15}}
\end{figure}

We show the FUV-based SFRs against the derived M$_{*}$ of the inner (open red triangles), outer regions (open blue squares), and the entire galaxies (open black circles) for the 27 GLSBGs in Figure ~\ref{fig:sfms_annuli_galaxyAll}, where the outer regions are systematically much closer to the MS than the outer regions which severely deviate from the MS. This tendency is more clearly revealed by the mean values of the inner (the filled red triangle), outer (the filled blue square) regions, and the entire galaxies (the filled black circle). It implies that the bulge-dominated components dominate the deviation of the entire GLSBGs from the MS due to the much lower SFRs of the bulges. However, the disk-dominated components are the dominant contributors to the global SFR of the entire GLSBGs. If only the disk-dominated components of these GLSBGs were adopted, the deviation from the MS would be largely mitigated, with some GLSBGs even approaching to follow the MS.

Furthermore, we zoom in on each individual GLSBG in Figures ~\ref{fig:sfms_annuli_galaxy12} and ~\ref{fig:sfms_annuli_galaxy15}.
In both figures, the inner part (red triangle) drops off from MS (E07 - black solid line; S16 - red dashed line) more obviously, while this drop is reduced in different degrees by the outer part (blue square) for each GLSBG. The inner part of the giant galaxy is conventionally dominated by the bulge component which has little contribution to star formation but occupies a considerable contribution to the stellar mass of the galaxy so the bulge should be the main driving force pulling the entire GLSBG off below the MS. In contrast, the outer part of the GLSBG is less affected by the central bulge component. Instead, it is mainly occupied by its large disk with relatively higher star formation but lower stellar mass than the bulge. So the deviation of the outer part from MS is reduced due to its enhanced star formation rate and decreased stellar mass. In some cases,  the outer part even tends to follow the MS. 

\begin{figure}[htbp]
\centering
\includegraphics[width=1.0\textwidth]{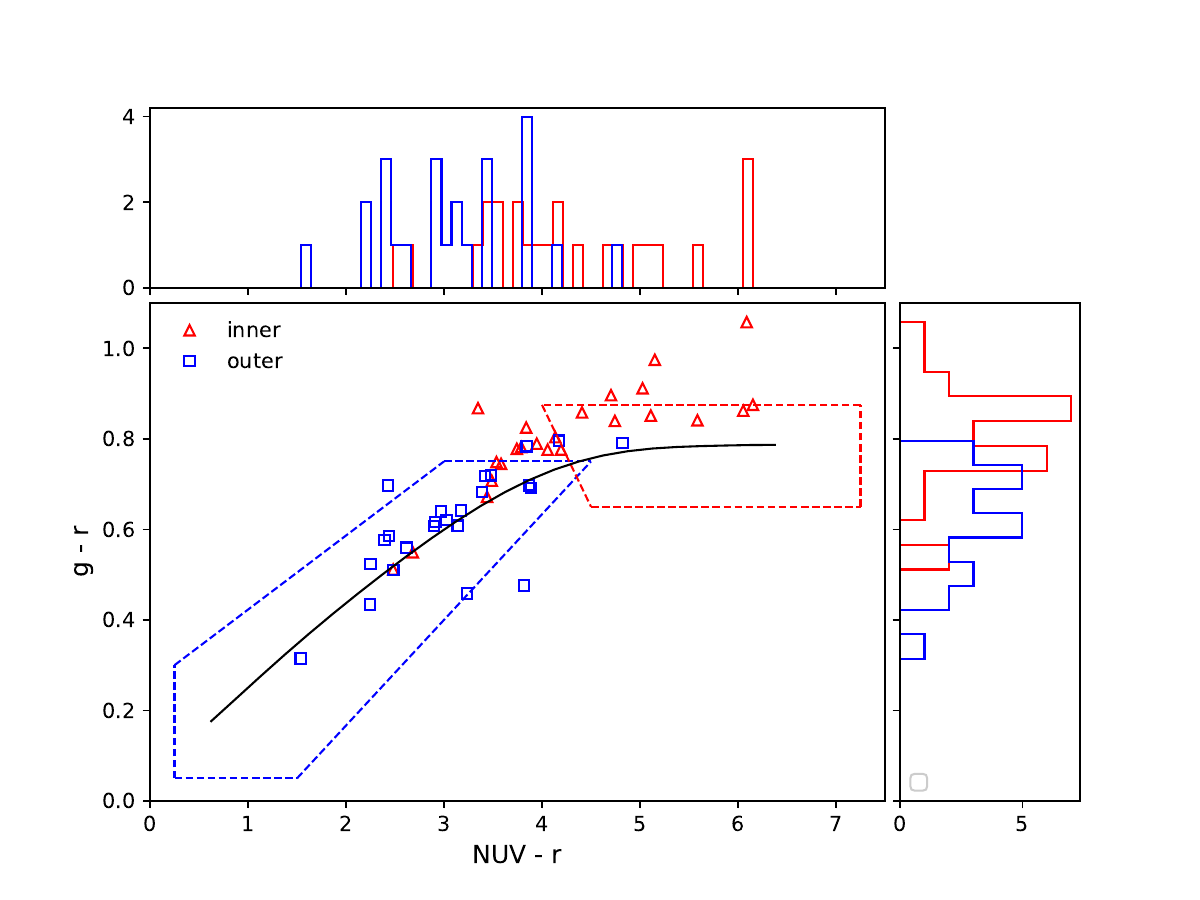}
\caption{The colors of NUV - $r$ and $g$ - $r$ of the inner (red triangle) and outer parts (blue square) for each GLSBG. From \citet{2012MNRAS.419.1727C}, it identifies the loci of blue cloud (blue dashed) and the red sequence (red dashed) based on SDSS galaxies, and also defines the relation between NUV -r and g - r for SDSS galaxies with r-band absolute magnitudes -23 $<$ M$_{r} <$ -21 mag (black solid curve). The right and top bars show the NUV - $r$ and $g$ - $r$ histogram distributions of the inner (red) and outer part (blue) of GLSBGs.  
\label{fig:color}}
\end{figure}

As is described in Section ~\ref{subsec:mstar-sfrs}, in order to make sure the measured fluxes on images of different bands are from the same region of a galaxy, optical images from DECaLs $g$ and $r$ bands are both lowered down in resolution to match the resolution of the GALEX UV image which is much lower. However, lowering the resolution flattens the optical surface brightness profile of the galaxy which spreads the effect of the central bulge out to the disk more, and makes it more difficult to completely decompose the bulge from the disk. We simply use R = R$_{e}$ as the demarcation line to divide a GLSBG into the inner and outer parts, only in order to make sure the inner part is absolutely occupied by the central bulge and the outer part is dominated by the disk in a large extent. It is evident in Figure ~\ref{fig:color}, the inner parts are indeed redder than the outer parts in general under this demarcation schema. We show the boundaries of blue clouds (blue dashed), red sequence (red dashed) identified for SDSS normal galaxies \citet{2012MNRAS.419.1727C}, and also the given relation between NUV - r  and g - r colors defined on SDSS normal galaxies with -23$<$M$_{r}<$-21 (black solid curve). Obviously, the outer parts (blue open squares) of GLSBGs behave similarly to the normal star-forming galaxies in the NUV - r  and g - r relation (black solid curve). The inner parts of GLSBGs appear to offset above from the relation, and the inner parts of some galaxies (UGC1752, UGC1455, Malin 1, UGC1922 with the offset from small to large) are even outside the upper boundary of the red sequence identified for normal SDSS galaxies. As is investigated in \citet{1998AJ....116.1650S}(S98), the majority of GLSBGs have a sizable bulge each that is outside the morphological class, including the four inner-part outliers (UGC1752, UGC1455, Malin 1, UGC1922) in our Figure ~\ref{fig:color}. This is likely a reason for the behavior of the inner parts of GLSBGs not following the relation between NUV - r and g - r at the redder end defined for normal galaxies. Besides, it is also investigated that half of the sample of GLSBGs in S98 have very low-ionization AGN behaviors, including UGC4422, UGC1455, UGC1922, UGC4219, and UGC6614 in our Figure ~\ref{fig:color}. Low-ionization AGNs should have low contribution to the optical light of the inner parts of the GLSBG hosts, so this might be also a reason for the behaviors of the inner parts of GLSBGs in the figure. 

As far as we can see from Figure ~\ref{fig:sfms_annuli_galaxy12} and ~\ref{fig:sfms_annuli_galaxy15}, such a demarcation schema work for most of the GLSBGs since the inner and outer parts could be clearly disentangled in the aspect of SFR and M$_{*}$, although the outer M$_{*}$ are not strikingly decomposed from the inner M$_{*}$ as expected due to the effect of the `flattened' bulge in the optical images (after the optical resolution is lowered down) on the outer part. For UM163 and UGC905, the demarcation schema does not work well as the outer M$_{*}$ surpasses the inner M$_{*}$ which implies that a larger radius threshold is required to better distinguish the disk-dominated part from the bulge-dominated part in the aspect of stellar mass. Albeit, in the aspect of SFR, the inner and outer parts could be distinguished in varying degrees for all GSLBGs, sufficiently demonstrating that the deviation of entire GLSBGs from the MS shape should be largely caused by the bulge components, and if only considering the pure diffuse disk, the deviation from the MS would be seriously reduced in the SFR-M$_{*}$ plane. This conclusion deduced from GLSBGs well supports that the turndown of the MS shape at high M$_{*}$ is mainly attributed by the growth of the central bulge that is quiescent with little SF but the disk component still tends to form stars at a consistent level with MS galaxies. 

For GLSBGs, the outer parts are highly distinguished from the inner parts in NUV - $r$ and $g$ - $r$ colors (Figure ~\ref{fig:color}). The inner parts are quite red while the outer parts are blue, indicating divergent origins of the two parts of GLSBGs. The stellar populations of the inner parts are relatively old and must have already formed and evolved, while the stellar populations of the outer parts are relatively young and star formation is still ongoing in the outer parts. This demonstrates an inside-out star formation mechanism for GLSBGs. Additionally, these GLSBGs show few morphological signs of merging and interactions with other galaxies (Figure ~\ref{fig:glsbg_set1} and ~\ref{fig:glsbg_set2}), which inclines us towards a picture that the large, extended low surface brightness disks of GLSBGs form and grow internally by themselves with little external effects after the central, bright bulge formed and evolved earlier. We plan to carry out detailed studies on the formation and evolution of GLSBGs with high resolution H{\sc{i}} data and optical long-slit spectra in our future work.  

\section{Summary}\label{sec:sum}
A sample of local GLSBGs observed by both DECaLs optical and GALEX UV imaging surveys is collected from the literature. The measurement of their stellar masses and FUV-based SFRs show that these GLSBGs have low SFRs and high M$_{*}$, deviating from the MS trajectory apparently at the high mass regime (M$_{*} >$ 10$^{10}$M$_{\odot}$) in the SFR-M$_{*}$ plane. Their sSFRs are lower than the characteristic value for the star-forming galaxies with M$_{*}$ = 10$^{11}$M$_{\odot}$ at z = 0 (sSFR $<$ 0.1 Gyr$^{-1}$). The deviation of the GLSBG as a whole from the MS should relate to the low molecular gas content and low SFE. These GLSBGs are H{\sc{i}}-rich systems with higher H{\sc{i}} gas mass fractions (f$_{\rm HI}$) than the S16 MS galaxies, but poor in H$_{2}$ because of quite a low efficiency of H{\sc{i}}-to-H$_{2}$ conversion. The possible reasons for the low transition are the lack of metal and dust, and the low H{\sc{i}} gas surface densities that are far below the minimum H{\sc{i}} surface density of 6 - 8 M$_{\odot}$ pc$^{-2}$ required for shielding H$_{2}$ against photodissociation.   

We discuss the contributions of the inner (bulge-dominated) and outer (disk-dominated) parts of GLSBGs to the entire deviations of GLSBGs from the MS. Generally, the bulge-dominated parts have lower SFRs and relatively higher M$_{*}$ than the disk-dominated parts of GLSBGs, and are the main force pulling the entire GLSBG off from the MS. However, the deviations from the MS are clearly reduced if the disk-dominated parts of the GLSBG are only considered in the SFR - M$_{*}$ plane.  For some cases, the disk-dominated parts even tend to follow the MS shape. In contrast to the red, inner, bulge-dominated parts, the outer, disk-dominated parts of GLSBGs are bluer and follow the NUV - r versus g - r relation for normal SFGs. The color difference between the inner and outer parts also implies an inside-out mechanism for the star formation of GLSBGs. 
They show few signs of tidal interactions, supporting that the large, extended low surface brightness disks are not likely to grow via recent major mergers.
\begin{acknowledgments}
We thank the referee for the helpful comments. D.W. is supported by the National Natural Science Foundation of China (NSFC) grant Nos. U1931109 and the Youth Innovation Promotion Association, CAS (No. 2020057). C.C. is supported by the NSFC No. 11803044, 11933003, 12173045 and also supported by the CAS South America Center for Astronomy (CASSACA). 
\end{acknowledgments}
%
\vspace{0mm}
\bibliography{SFMS_GLSBG}{}
\bibliographystyle{aasjournal}
\end{document}